\documentclass[twocolumn,superscriptaddress]{revtex4-1}
\usepackage{graphicx}
\usepackage{units}
\usepackage{multirow}
\usepackage{bigstrut}
\usepackage{overpic}
\usepackage{array}
\usepackage{color}
\usepackage{amsmath}
\usepackage{xspace}
\usepackage{amsfonts}
\usepackage{subfigure}
\usepackage{verbatim}
\usepackage{footmisc}
\usepackage{color}
\usepackage{epstopdf}
\RequirePackage{lineno}
\usepackage[colorlinks,linkcolor=blue,anchorcolor=blue,citecolor=blue]{hyperref}
\newcommand{\pip}{\pi^+}
\newcommand{\pizero}{\pi^0}
\newcommand{\pim}{\pi^-}

\newcommand{\mev}{\,\unit{MeV}}

\newcommand{\mevcc}{\,\unit{MeV}/c^2}

\newcommand{\jpsi}{J/\psi}
\newcommand{\Lambdabar}{\bar{\Lambda}}

\begin{document}%%

%\setpagewiselinenumbers
%\modulolinenumbers[2]
%\linenumbers

\title{\bf\boldmath Search for invisible decays of the $\Lambda$ baryon}
\date{\it \small \bf \today}
%Created at 2021-08-11
\author{\small
M.~Ablikim$^{1}$, M.~N.~Achasov$^{10,b}$, P.~Adlarson$^{66}$, S. ~Ahmed$^{14}$, M.~Albrecht$^{4}$, R.~Aliberti$^{27}$, A.~Amoroso$^{65A,65C}$, M.~R.~An$^{31}$, Q.~An$^{62,48}$, X.~H.~Bai$^{56}$
% ...
\\
\vspace{0.2cm}
(BESIII Collaboration)\\
\vspace{0.2cm} {\it
$^{1}$ Institute of High Energy Physics, Beijing 100049, People's Republic of China\\
% ...
$^{4}$ Bochum Ruhr-University, D-44780 Bochum, Germany\\
% ...
$^{10}$ G.I. Budker Institute of Nuclear Physics SB RAS (BINP), Novosibirsk 630090, Russia\\
% ...
$^{14}$ Helmholtz Institute Mainz, Staudinger Weg 18, D-55099 Mainz, Germany\\
% ...
$^{27}$ Johannes Gutenberg University of Mainz, Johann-Joachim-Becher-Weg 45, D-55099 Mainz, Germany\\
% ...
$^{31}$ Liaoning Normal University, Dalian 116029, People's Republic of China\\
% ...
$^{48}$ State Key Laboratory of Particle Detection and Electronics, Beijing 100049, Hefei 230026, People's Republic of China\\
% ...
$^{56}$ University of Jinan, Jinan 250022, People's Republic of China\\
% ...
$^{62}$ University of Science and Technology of China, Hefei 230026, People's Republic of China\\
% ...
$^{65}$ University of Turin and INFN, (A)University of Turin, I-10125, Turin, Italy; (B)University of Eastern Piedmont, I-15121, Alessandria, Italy; (C)INFN, I-10125, Turin, Italy\\
$^{66}$ Uppsala University, Box 516, SE-75120 Uppsala, Sweden\\
% ...
$^{b}$ Also at the Novosibirsk State University, Novosibirsk, 630090, Russia\\
% ...
}\vspace{0.4cm}}

\vspace{4cm}
\begin{abstract}
A search for invisible decays of the $\Lambda$ baryon is carried out in the process $\jpsi\to\Lambda\bar{\Lambda}$ based on $(1.0087\pm0.0044)\times10^{10}$ $J/\psi$ events collected with the BESIII detector located at the BEPCII storage ring. No signals are found for the invisible decays of $\Lambda$ baryon, and the upper limit of the branching fraction is determined to be $7.4 \times 10^{-5}$ at the 90\% confidence level. 
This is the first search for invisible decays of baryons; such searches will play an important role in constraining dark sector models related to the baryon asymmetry.
\end{abstract}

\maketitle
%%%%%%%%%%%%%%%%%%%%%%%%%%%%%%%%%%%%%%%%%%%%%%%%%%%%%%%%%%%%%%%%
%%%%%     Introduction       Part                  %%%%%%%%%%%%%
%%%%%%%%%%%%%%%%%%%%%%%%%%%%%%%%%%%%%%%%%%%%%%%%%%%%%%%%%%%%%%%%
%\begin{multicols}{2}

Understanding dark matter is a highly topical subject in both astronomy and particle physics. Although strong indirect evidence for the existence of dark matter is obtained via astronomy, there is no direct evidence from collider experiments yet. 
On the other hand, the asymmetry between matter and antimatter in the universe indicates that baryon number ($B$) conservation  is violated~\cite{Sakharov:1967dj}. 
The baryon matter density and the dark matter density are similar, $\rho_{\rm DM } \approx 5.4~\rho_{\rm  baryon}$, which may hint at a common origin of these two unsolved questions. 
Dark matter may be represented by baryon matter with invisible final state~\cite{Alonso-Alvarez:2021oaj}.
In the asymmetric dark matter scenario \cite{Nussinov:1985xr, Dodelson:1989cq, Barr:1990ca, Kaplan:1991ah, Farrar:2005zd, Kaplan:2009ag}, the dark matter and baryon asymmetry puzzles may be related and the dark matter mass could be in the order of GeV. 
Simultaneous generation of the necessary baryon asymmetry and dark matter density is possible \cite{Shelton:2010ta, Davoudiasl:2010am, Gu:2010ft}. 
Those models usually contain a neutron portal operator $u^c d^c d^c$ \cite{Petraki:2013wwa, Zurek:2013wia}, where $q^c$ is the right-handed quark, to couple directly with dark matter or dark sector particles via effective operators obtained after integrating out a color-triplet scalar. 
Such operators can generally introduce $B$ violation to the Standard Model sector. 
This type of interaction has been used to explain the discrepancy of neutron lifetime measurements in the beam method~\cite{Bondarenko:1978dn,Byrne:1996zz,Yue:2013qrc,Bottyan:2011wv} and the bottle method~\cite{Mampe:1993an,Serebrov:2004zf,Pichlmaier:2010zz,Steyerl:2012zz,Arzumanov:2015tea,Serebrov:2017bzo,Pattie:2017vsj, Ezhov:2014tna}, by requiring $1\%$ of the neutrons to decay into dark matter particles \cite{Fornal:2018eol}. 
Neutrons and antineutrons can also oscillate if the neutron portal couples to a Majorana dark sector particle, leading to a $\Delta B = 2$ process~\cite{Phillips:2014fgb}.
A mirror world with mirror dark matter can lead to a similar phenomenon~\cite{TAN2019134921,Tan:2020gpd}.
Recently, exotic baryon number violating decays of hydrogen atoms to dark sector particles and neutrinos through the neutron portal have been discussed~\cite{McKeen:2020vpf}. 
Moreover, allowing for quarks from the second and third generation in the neutron portal will generally result in new portals, which can lead to heavy flavor meson and baryon decays into dark sector particles.
Specifically, $b$ hadrons like $B^0_{d,s}$, $B^\pm$ and $\Lambda_b$ have been discussed~\cite{Elor:2018twp, Alonso-Alvarez:2021qfd}; their exotic decays could be probed at Belle II and in the LHC experiments.
As many theories and experiments suggest a potential correlation between baryon symmetry and dark sector, study of baryon invisible decays is well motivated.

The search for invisible decays of neutral hadrons is highly interesting, 
since such decays could involve a potential dark matter candidate. 
Since the missing energy cannot be fully measured at hadron colliders~\cite{Borsato:2021aum}, such studies are difficult to perform there.  
Electron-positron collision experiments such as BESIII and Belle II have the ability to probe invisible decays, benefiting from a well-defined production process and a clean reaction environment.
Stringent limits on the invisible decays of $\Upsilon$~\cite{Aubert:2009ae}, $J/\psi$~\cite{Ablikim:2007ek}, $B^{0}$~\cite{Hsu:2012uh}, ${{\mathit \eta}^{(\prime)}}$~\cite{Ablikim:2012gf}, $\pi^{0}$~\cite{NA62:2020pwi}, $D^{0}$~\cite{Lai:2016uvj}, $\omega$~\cite{Ablikim:2018liz}, and $\phi$~\cite{Ablikim:2018liz}, mesons have already been determined by several experiments. However, no experimental study of invisible baryon decays has been carried out yet. 
A search for invisible decays of the $\Lambda$ hyperon may provide 
information which can help to understand invisible decays of neutrons. 

In this paper, the first experimental search for invisible decays of the $\Lambda$ baryon is carried out using $(1.0087\pm0.0044)\times10^{10}$ $\jpsi$ events~\cite{JpsiNumber} accumulated at the center-of-mass energy $\sqrt{s}=3.097$ GeV with the BESIII detector~\cite{Ablikim:2009aa} at the BEPCII storage ring~\cite{Ablikim:2019hff,Yu:IPAC2016-TUYA01}. Taking advantage of the clean environment of $\Lambda\bar{\Lambda}$ pairs produced in $\jpsi\to\Lambda\bar{\Lambda}$, one $\Lambdabar$ is explicitly reconstructed via $\Lambdabar\to \bar{p} \pi^+$, allowing to search for invisible decays of the recoiling $\Lambda$. 
Invisible $\Lambdabar$ decays are not pursued in this work, because the dominant background from $\Lambdabar\to\bar{n}\pi^0$ is hard to estimate due to the difficulties in simulating the hadronic interactions of antineutrons with the detector material. An antineutron can induce a large number of showers spread over the detector, which introduces difficulties of finding a clean antineutron control sample to correct the Monte Carlo (MC) simulation.

%%%%%%%%%%%%%%%%%%%%%%%%%%%%%%%%%%%%%%%%%%%%%%%%%%%%%%%%%%%%%%%%
%%%%%     Detector and software Part               %%%%%%%%%%%%%
%%%%%%%%%%%%%%%%%%%%%%%%%%%%%%%%%%%%%%%%%%%%%%%%%%%%%%%%%%%%%%%%

The BESIII detector records symmetric $e^+e^-$ collisions provided by the BEPCII storage ring. The cylindrical core of the BESIII detector covers nearly 93\% of the full solid angle and consists of a helium-based multilayer drift chamber~(MDC), a plastic scintillator time-of-flight system~(TOF), and a CsI(Tl) electromagnetic calorimeter~(EMC), which are all enclosed in a superconducting solenoidal magnet providing a 1.0~T (0.9~T in 2012) magnetic field. About 91\% of the data was taken with the larger field.  The solenoid is supported by an octagonal flux-return yoke with resistive plate counter muon identification modules interleaved with steel. The charged-particle momentum resolution at $1~{\rm GeV}/c$ is $0.5\%$. The EMC measures energies of photons with a resolution of $2.5\%$ ($5\%$) at $1$~GeV in the barrel (end cap) region. The time resolution in the TOF barrel region is 68~ps, and it was 110~ps in the end cap region before 2015. An end cap TOF system upgrade in 2015 using multigap resistive plate chamber technology improved the time resolution to 60~ps~\cite{etof} 
for about 87\% of the dataset.  

Simulated data samples produced with a {\sc geant4}-based~\cite{geant4} MC package, which includes the geometric description and response of the detector, are used to determine the detection efficiency and understand background contributions. The simulation takes into account the beam energy spread and initial state radiation in the $e^+e^-$ annihilations with the generator {\sc kkmc}~\cite{ref:kkmc}. A sample of $1.0011\times10^{10}$ simulated inclusive $\jpsi$ decays is used. The inclusive MC sample includes both the production of the $J/\psi$ resonance and the continuum processes incorporated in {\sc kkmc}. The known decays are modeled with {\sc evtgen}~\cite{ref:evtgen} using branching fractions taken from the Particle Data Group~\cite{pdg}. The remaining unknown $J/\psi$ decays are modeled with {\sc lundcharm}~\cite{ref:lundcharm}. Final state radiation of charged final state particles is incorporated using the {\sc photos} package~\cite{photos}.
For the signal MC sample, the helicity formalism~\cite{Ablikim:2018zay} is applied for $J/\psi\to\Lambda \Lambdabar$ followed by the $\Lambda$ decaying invisibly and the $\Lambdabar$ decaying into $\bar{p}\pi^{+}$.

%%%%%%%%%%%%%%%%%%%%%%%%%%%%%%%%%%%%%%%%%%%%%%%%%%%%%%%%%%%%%%%%
%%%%%            Analysis Strategy.                 %%%%%%%%%%%%
%%%%%%%%%%%%%%%%%%%%%%%%%%%%%%%%%%%%%%%%%%%%%%%%%%%%%%%%%%%%%%%%
In the two-body decay of $\jpsi\to\Lambda\bar{\Lambda}$, the $\Lambda$ can be inferred by detecting the $\Lambdabar$ decay, which provides a model-independent way to study $\Lambda$ decays without relying on the total number of $J/\psi$ decays. In this philosophy, the number of events with a $\Lambdabar$ detected is denoted as $N_{\rm tag}$, and the number of events in which the signal decay of the accompanying $\Lambda$ is also detected is denoted as $N_{\rm sig}$.
 Therefore, the branching fraction for the signal decay is given as
\begin{eqnarray}
 \mathcal{B}(\Lambda\to \rm invisible )=\frac{\it N_{\mathrm{sig}}} {\it N_{\mathrm{tag}} \cdot (\varepsilon_{\mathrm{sig}}/\varepsilon_{\mathrm{tag}})}
\label{Eq:BF}
\end{eqnarray}
Here, $\varepsilon_{\mathrm{sig}}$ is the detection efficiency when both $\Lambdabar$ and $\Lambda$  decays are selected and $\varepsilon_{\mathrm{tag}}$ is the efficiency of only detecting the $\Lambdabar$ in $\jpsi\to\Lambda\bar{\Lambda}$. A semiblind analysis is performed to avoid possible bias, where 1\% of the full data sample is used to validate the analysis strategy. The final results are obtained with the full data sample only after the analysis method is frozen.

%%%%%%%%%%%%%%%%%%%%%%%%%%%%%%%%%%%%%%%%%%%%%%%%%%%%%%%%%%%%%%%%
%%%%%  Single-tag event selection and yields        %%%%%%%%%%%%
%%%%%%%%%%%%%%%%%%%%%%%%%%%%%%%%%%%%%%%%%%%%%%%%%%%%%%%%%%%%%%%%
To select $\Lambdabar$ tag events, at least two charged particle tracks must be reconstructed in the  MDC.  They are required to have a polar angle $\theta$ with respect to the symmetry axis of the MDC satisfying $|\cos\theta|<0.93$.   Particle identification uses measurements of d$E$/d$x$ in the MDC, forming likelihoods $\mathcal{L}(h)~(h=K,\pi, p)$ for each hadron $h$ hypothesis. The $\bar{p}$ is identified by requiring $\mathcal{L}(\bar{p})>\mathcal{L}(\pim)$ and $\mathcal{L}(\bar{p})>\mathcal{L}(K^-)$. The $\pi^+$ is identified by requiring $\mathcal{L}(\pip)>\mathcal{L}(p)$ and $\mathcal{L}(\pip)>\mathcal{L}(K^+)$. At least one $\pip$ and one $\bar{p}$ are required. 

To reconstruct the $\bar{\Lambda}$,  the $\bar{p}\pip$ tracks are constrained to a common vertex by applying a vertex fit, and the corresponding $\chi^2$ is required to be less than 200. To further suppress background contributions, a constraint of the $\bar{\Lambda}$ momentum vector pointing to the interaction point is implemented and the corresponding $\chi^2$ must be less than 10.
The fitted decay length is required to be larger than three times its resolution. The updated momentum of the $\bar{p}\pip$ pairs after the constrained fit is used in the subsequent analysis. In addition, the cosine of the polar angle of the $\Lambdabar$ candidate is required to be less than 0.70 to ensure that the $\Lambda$ decay products are in the acceptance of the EMC barrel region which covers polar angles of $|\cos\theta|<0.83$.
To obtain a good resolution of the event start time and hence ensure that all the showers are in time with the event, one of the two charged tracks is required to leave cluster information in one of the TOF layers.  The relative efficiency of this requirement is $(89.67\pm0.15)\%$. 
The invariant mass of $\bar{p}\pi^{+}$ is required to be within 5 $\mevcc$ of the nominal $\bar{\Lambda}$ mass. Those events with exactly one $\bar{\Lambda}$ candidate are retained for further studies. 

\begin{figure}[h!]
\centering
\begin{overpic}[width=0.9\linewidth]{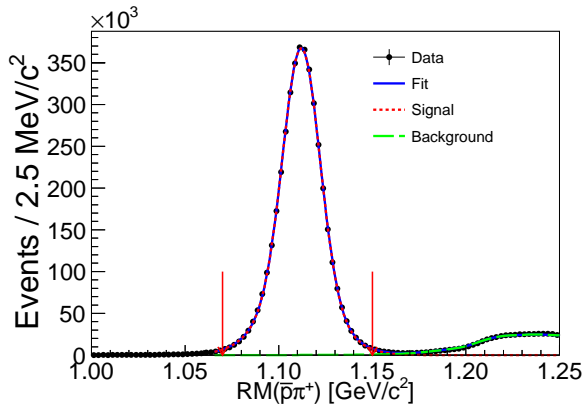}
\end{overpic}
\caption{Fit to the recoil mass of $\bar{p}\pip$. The black dots with uncertainties represent data and the blue solid line shows the total fit. The red curve and green long-dashed curve are the fitted signal and background contributions, respectively.}
\label{fig:fit_ST}
\end{figure}

The $\Lambdabar$ single tag yield $N_{\mathrm{tag}}$ is obtained from a binned maximum likelihood fit to the recoil mass distribution $RM(\bar{p}\pi^{+})$ as depicted in Fig.~\ref{fig:fit_ST}.
In the fit, the signal is modeled with a double Gaussian function, while background contributions are described with the shape derived from the inclusive MC sample.
We find $N_{\mathrm{tag}}=(4154428\pm2040)$ by integrating the signal function within $\pm40$ $\mevcc$ of the nominal $\Lambda$ mass. 
The signal purity is 99.85\% and the tagging efficiency is estimated to be $\varepsilon_{\mathrm{tag}}=(32.11\pm0.01)\%$ based on a MC sample of $J/\psi\to\Lambda \bar{\Lambda}$ with $\bar{\Lambda}\to \bar{p}\pi^+$ and inclusive $\Lambda$ decays. 

%%%%%%%%%%%%%%%%%%%%%%%%%%%%%
%%%%%%%%%%%%%%%%%%%%%%%%%%%%%

%%%%%%%%%%%%%%%%%%%%%%%%%%%%%%%%%%%%%%%%%%%%%%%%%%%%%%%%%%%%%%%%
%%%%%  Double-tag event selection                   %%%%%%%%%%%%
%%%%%%%%%%%%%%%%%%%%%%%%%%%%%%%%%%%%%%%%%%%%%%%%%%%%%%%%%%%%%%%%

To select signal candidates of the invisible $\Lambda$ decays, $RM(\bar{p}\pi^{+})$ is required to be within 40 $\mevcc$ of the nominal $\Lambda$ mass. 
It is required that there is no additional charged track; this has a relative efficiency of  $\varepsilon_{\mathrm{sig}}/\varepsilon_{\mathrm{tag}}=(94.64\pm0.17)\%$. 
The efficiency loss is mainly due to secondary tracks originating from the antiprotons interacting with the detector material. 
As the invisible $\Lambda$ decay final states do not deposit any energy in the EMC, the sum of energies of all the EMC showers not associated with any charged tracks, $E_{\rm EMC}$, can be used as a discriminator. 
Suppression of EMC showers from charged tracks is achieved by an isolation requirement: the angles between any shower included in the sum and the momenta of the $\pip$ and $\bar{p}$ tracks must be greater than $10^{\circ}$  and $20^{\circ}$, respectively. 

\begin{figure}[h!]
\centering
\begin{overpic}[width=0.9\linewidth]{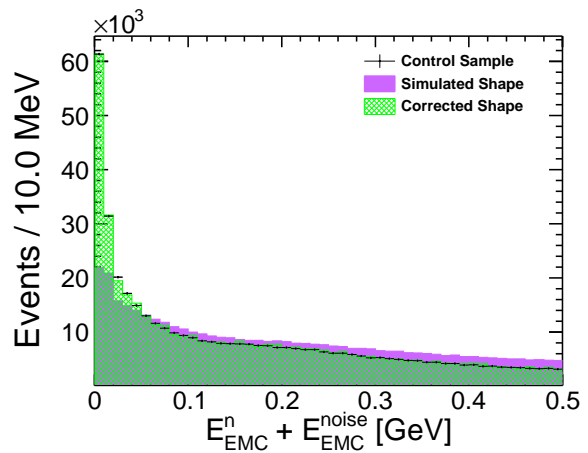}
\end{overpic}
\caption{The $E_{\rm EMC}^{n} + E_{\rm EMC}^{\rm noise}$ distribution in $\jpsi\to\bar{p}\pi^{+}n$ control sample. Data points with error bars represent neutron control sample in data. The purple full filled histogram shows the {\sc geant4}-based simulated shape, and the green cross filled histogram shows the corrected shape.}
\label{fig:EMCCS}
\end{figure}

As inferred from studies of the inclusive MC sample, $\Lambda\to n\pi^{0}$ is the dominant background after applying all selection conditions. 
However, due to the inaccurately modeled neutron interactions in the detector material by the {\sc geant4} package, the simulation of the energy deposits of neutrons in the EMC is unreliable, as illustrated in Fig.~\ref{fig:EMCCS}.
Therefore, the MC-derived shape of this background contribution is corrected based on a neutron control sample, which can be used to model the energy deposit in the EMC from the penetrating neutron. The energy deposit in the EMC can be divided into three parts, as detailed below
\begin{eqnarray}
E_{\rm EMC} = E_{\rm EMC}^{\pi^{0}} + E_{\rm EMC}^{n} + E_{\rm EMC}^{\rm noise} ,
\end{eqnarray}
where $E_{\rm EMC}^{\pi^{0}}$ is the energy due to electromagnetic showers from $\pizero$ decays, $E_{\rm EMC}^{n}$ is the energy deposited by neutrons, and $E_{\rm EMC}^{\rm noise}$ is the energy of showers unrelated to the event.
Among them, $E_{\rm EMC}^{\pi^{0}}$ is retained based on MC simulations as the interactions of photons or electrons with material are reliably described in the simulation. 
The information of  $E_{\rm EMC}^{\pi^{0}}$ is recorded by switching off the interactions of $\bar{p}, \pi^{+}$ and $n$ with the detector material in the exclusive simulation of the background process of $\jpsi\to\Lambda(n\pi^{0})\Lambdabar{(\bar{p}\pi^{+})}$.
For the sum  of $E_{\rm EMC}^{n}$  and $E_{\rm EMC}^{\rm noise}$, a neutron control sample of $\jpsi\to\Lambda(n\pi^{0})\Lambdabar{(\bar{p}\pi^{+})}$, $\pi^0\to\gamma\gamma$ is selected, where all final state particles are detected except the neutron. 
A series of shape templates as a function of the momentum and polar angle of the neutron is derived by summing up the energies of all the showers except the photon showers from $\pi^0$ decays, which are not associated with the $\bar{p}\pip$ tracks.
The corrected $E_{\rm EMC}$ shape for $\Lambda\to n \pizero$ background contributions is derived by combining $E_{\rm EMC}^{\pi^{0}}$ with a random value of the sum of $E_{\rm EMC}^{n}$ and $E_{\rm EMC}^{\rm noise}$ from the shape template according to the momentum and polar angle of the neutron in the exclusive MC simulations. 
The above mentioned correction procedure has been validated based on the control sample of $\jpsi\to p\bar{p}\pi^0$ and $\bar{p}\pi^{+}n$, where $p\bar{p}\pi^0$ is used to check the $\pizero$ simulation and $\bar{p}\pi^{+}n$ to cross check the two-dimensional resampling method on the EMC shower energy of the neutron control sample.
Figure~\ref{fig:EMCCS} shows that the resampled shape of $E_{\rm EMC}^{n}+E_{\rm EMC}^{\rm noise}$ well reproduce the original shape in $\jpsi\to\bar{p}\pi^{+}n$ control sample.

%%%%%%%%%%%%%%%%%%%%%%%%%%%%%
%%%%%%%%%%%%%%%%%%%%%%%%%%%%%

\begin{figure}[h!]
\centering
\begin{overpic}[width=0.9\linewidth]{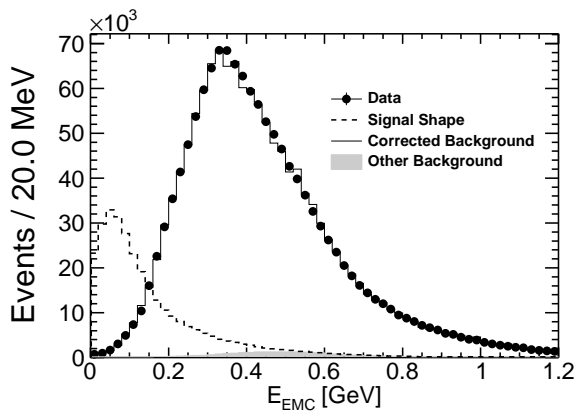}
\end{overpic}
\caption{The $E_{\rm EMC}$ distribution. The dots with uncertainties represent data. The dashed line shows the signal shape, where the corresponding yield is normalized arbitrarily for clarity. The solid line shows the $\Lambda\to n\pi^0$ background shape including the correction. The grey filled area shows the other background contributions mentioned in the text.}
\label{fig:EMC}
\end{figure}

The corrected distribution of $E_{\rm EMC}$ is shown in Fig.~\ref{fig:EMC}, where the resulting $E_{\rm EMC}$ distribution for the $\Lambda\to n\pi^{0}$ background and for other remaining minor $\Lambda$ decay background contributions, such as $\Lambda\to n\gamma$ with size of 0.5\% of $\Lambda\to n\pi^{0}$ background, in the inclusive MC sample is overlaid. 
The signal of invisible $\Lambda$ decays is expected to peak close to zero, as the dashed line in Fig.~\ref{fig:EMC} shows. 
The signal shape can be modeled by MC simulations from noise plus a small contribution from charged-particle showers leaking through the isolation requirement.

The total $E_{\rm EMC}$ distribution from the simulated $\jpsi$ decay background is found to agree well with the data, no obvious signal is observed. An upper limit (UL) at 90$\%$ confidence level (C.L.) on the branching fraction of invisible $\Lambda$ decays, $\mathcal{B}(\Lambda\to \rm invisible)$, is evaluated by a binned maximum likelihood fit to the $E_{\rm EMC}$ distribution, after taking into account the effects of statistical and systematic uncertainties. Here, the number of signal events ${\it N_{\mathrm{sig}}}$ is obtained by integrating the signal fraction of the fit function.

%%%%%%%%%%%%%%%%%%%%%%%%%%%%%
%%%%%%%%%%%%%%%%%%%%%%%%%%%%%

%%%%%%%%%%%%%%%%%%%%%%%%%%%%%%%%%%%%%%%%%%%%%%%%%%%%%%%%%%%%%%%%
%%%%%  Systematic uncertainty                       %%%%%%%%%%%%
%%%%%%%%%%%%%%%%%%%%%%%%%%%%%%%%%%%%%%%%%%%%%%%%%%%%%%%%%%%%%%%%

The systematic uncertainties on $\mathcal{B}(\Lambda\to \rm invisible)$ due to the selection criteria used for tagging $\Lambdabar$ candidates cancel in Eq.~(\ref{Eq:BF}). The systematic uncertainty related to the determination of $N_{\rm tag}$ is found to be negligible. The sources of the dominant systematic uncertainties are the selection condition regarding the separation angle between the EMC shower and the antiproton, the choice of binning in fitting to the $E_{\rm EMC}$ distributions and the requirement of no additional charged track. In the former two cases, we vary the condition (the binning) as summarized in Table~\ref{sys_source}.

\begin{table}[h]
\begin{center}
	\caption{Sources of systematic uncertainties, which are taken into account in estimating the UL of the invisible $\Lambda$ decay rate.}\label{sys_source}
	\begin{tabular}{l|c}\hline\hline
		Source & Choice or uncertainty \\\hline
		Shower separation angle&  $18^{\circ}$, $20^{\circ}$ and $22^{\circ}$ \\
		Bin width &  10, 20, 30, 40, 50 $\mev$ \\ \hline
		No additional charged track & $0.6\%$\\
		\hline\hline
		\end{tabular}
\end{center}
\end{table}

Antiproton interactions with the EMC material can produce a large number of fake photon signals over a large area. The requirement on the opening angle between the shower direction and the $\bar{p}$ track extrapolated to the EMC can affect the $E_{\rm EMC}$ shape in both signal and background processes. To take this effect into account, different choices of separation angles are tested in the analysis procedure. In the likelihood fit, five different choices of bin widths are considered with a minimum width of 10$\mev$. 
The requirement of no additional charged tracks affects the estimation of the detection efficiency and the $E_{\rm EMC}$ distribution. To study this systematic effect, a clean control sample of $\jpsi \to \Lambda(p\pi^{-}) \bar{\Lambda}(\bar{p}\pi^{+})$ is selected requiring a full detection of all final state charged tracks. The efficiency difference of the further requirement of no additional charged tracks between data and MC simulations is found to be 0.6$\%$.
%%%%%%%%%%%%%%%%%%%%%%%%%%%%%%%%%%%%%%%%%%%%%%%%%%%%%%%%%%%%%%%%
%%%%%  UL estimation                                %%%%%%%%%%%%
%%%%%%%%%%%%%%%%%%%%%%%%%%%%%%%%%%%%%%%%%%%%%%%%%%%%%%%%%%%%%%%%

\begin{figure}[htb!]
\centering
\begin{overpic}[width=0.9\linewidth]{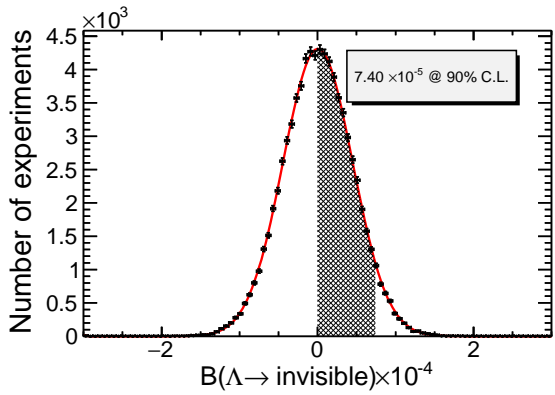}
\end{overpic}
\caption{Distribution of the estimated $\mathcal{B}(\Lambda\to \rm invisible)$ in pseudosamples. The shaded area corresponds to the 90\% coverage in the physical region.}
\label{fig:UL}
\end{figure}

A modified frequentist approach~\cite{Ablikim:2015djc,Ablikim:2017twd}, which incorporates all the systematic and statistical uncertainties, is adopted to estimate the UL of $\mathcal{B}(\Lambda\to \rm invisible)$.  
In the procedure, thousands of pseudodata samples are generated according to the $E_{\rm EMC}$ distributions in data. In each sample, the number of events is randomly chosen with a Poisson distribution with a mean value corresponding to the data. This approach incorporates the statistical fluctuation. A binned maximum likelihood fit is implemented, where the signal model is taken from the signal MC sample, and $\Lambda\to n\pizero$ background contributions are described with the MC-determined shape after correction with the neutron control sample.  The sizes of these two components are free parameters in each fit.  
The size and shape of the remaining $\Lambda$ decay backgrounds are fixed according to the inclusive MC sample; this is found to cause negligible uncertainty.  
Values of bin width and the shower separation angle, as listed in Table~\ref{sys_source}, are randomly selected and applied to the corresponding signal and background models.  To calculate $\mathcal{B}(\Lambda\to \rm invisible)$, the involved efficiency ratio $\varepsilon_{\mathrm{sig}}/\varepsilon_{\mathrm{tag}}$ in Eq.~(\ref{Eq:BF}) is scaled with a factor randomly sampled from a Gaussian function with mean of 1 and width of 0.006.
The resulting distribution of the calculated branching fractions in the pseudosamples is shown in Fig.~\ref{fig:UL}, described by a Gaussian function. The UL on $\mathcal{B}(\Lambda\to \rm{invisible})$ at 90\% C.L. is determined by integrating in the physical region ($\mathcal{B}>0$). The corresponding UL is obtained to be $7.4\times10^{-5}$, including the systematic uncertainties.

%%%%%%%%%%%%%%%%%%%%%%%%%%%%%%%%%%%%%%%%%%%%%%%%%%%%%%%%%%%%%%%%
%%%%%  Summary.                                     %%%%%%%%%%%%
%%%%%%%%%%%%%%%%%%%%%%%%%%%%%%%%%%%%%%%%%%%%%%%%%%%%%%%%%%%%%%%%

In summary, with a sample of $(1.0087\pm0.0044)\times10^{10}$ $J/ \psi$ events collected by the BESIII detector, the first search for invisible decays of the $\Lambda$ hyperon is carried out. 
This is the first direct experimental search for invisible decays of baryons. 
No obvious signal is observed and the UL on the decay rate is $\mathcal{B}(\Lambda\to \rm invisible )<7.4\times10^{-5}$ at 90$\%$ C.L, which is consistent with the prediction of $4.4\times10^{-7}$ from the mirror model~\cite{Tan:2020gpd}. 
This result sheds light on the neutron lifetime measurement puzzle~\cite{Bondarenko:1978dn,Byrne:1996zz,Yue:2013qrc,Bottyan:2011wv,Mampe:1993an,Serebrov:2004zf,Pichlmaier:2010zz,Steyerl:2012zz,Arzumanov:2015tea,Serebrov:2017bzo,Pattie:2017vsj, Ezhov:2014tna} and helps to constrain dark sector models related to the baryon asymmetry.
%This result sheds light on further understanding the puzzling of neutron lifetime~\cite{Bondarenko:1978dn,Byrne:1996zz,Yue:2013qrc,Bottyan:2011wv,Mampe:1993an,Serebrov:2004zf,Pichlmaier:2010zz,Steyerl:2012zz,Arzumanov:2015tea,Serebrov:2017bzo,Pattie:2017vsj, Ezhov:2014tna} and helps to constrain dark sector models related to the baryon asymmetry.

%\section{\boldmath Acknowledgments}
%\acknowledgments

The BESIII collaboration thanks the staff of BEPCII and the IHEP computing center for their strong support. We thank Yi Liao, Jia Liu and Fu-Sheng Yu for useful discussions. This work is supported in part by National Key Research and Development Program of China under Contracts Nos. 2020YFA0406400, 2020YFA0406300; National Natural Science Foundation of China (NSFC) under Contracts Nos. 11625523, 11635010, 11735014, 11822506, 11835012, 11935015, 11935016, 11935018, 11961141012, 12022510, 12025502, 12035009, 12035013, 12061131003; the Chinese Academy of Sciences (CAS) Large-Scale Scientific Facility Program; Joint Large-Scale Scientific Facility Funds of the NSFC and CAS under Contracts Nos. U1732263, U1832207; CAS Key Research Program of Frontier Sciences under Contract No. QYZDJ-SSW-SLH040; 100 Talents Program of CAS; INPAC and Shanghai Key Laboratory for Particle Physics and Cosmology; ERC under Contract No. 758462; European Union Horizon 2020 research and innovation programme under Contract No. Marie Sklodowska-Curie grant agreement No 894790; German Research Foundation DFG under Contracts Nos. 443159800, Collaborative Research Center CRC 1044, FOR 2359, GRK 214; Istituto Nazionale di Fisica Nucleare, Italy; Ministry of Development of Turkey under Contract No. DPT2006K-120470; National Science and Technology fund; Olle Engkvist Foundation under Contract No. 200-0605; STFC (United Kingdom); The Knut and Alice Wallenberg Foundation (Sweden) under Contract No. 2016.0157; The Royal Society, UK under Contracts Nos. DH140054, DH160214; The Swedish Research Council; U. S. Department of Energy under Contracts Nos. DE-FG02-05ER41374, DE-SC-0012069.

\end{document}